\documentclass{article}

\usepackage{arxiv}

\usepackage[utf8]{inputenc} 
\usepackage[T1]{fontenc}    
\usepackage{hyperref}       
\usepackage{url}            
\usepackage{booktabs}       
\usepackage{nicefrac}       
\usepackage{microtype}      
\usepackage{graphicx}
\usepackage{natbib}
\usepackage{doi}

\usepackage[thicklines]{cancel}
\usepackage{algorithmic}
\usepackage{amsmath,amssymb,amsfonts,bbm}
\usepackage{cleveref} 
\usepackage{comment}
\usepackage{textcomp}
\usepackage{multirow}
\usepackage{subcaption}
\usepackage{wrapfig}
\usepackage{epstopdf}
\usepackage[table]{xcolor}
\usepackage{adjustbox}

\usepackage{draftwatermark}
\SetWatermarkText{DRAFT}
\SetWatermarkScale{1}

\newcommand{\myccO}{\cellcolor{myOrange}}

\newcommand{\myccLC}{\cellcolor{LightCyan}}

\definecolor{LightCyan}{rgb}{0.88,1,1}
\definecolor{myOrange}{HTML}{FFA07A} 
\definecolor{myBlue}{HTML}{728ed6} 
\definecolor{darkgreen}{HTML}{037D50}

\title{Emotion Classification from Multi-Channel EEG Signals Using HiSTN: A Hierarchical Graph-based Spatial-Temporal Approach}

\author{ \href{https://orcid.org/0000-0002-4862-7182}{\includegraphics[scale=0.06]{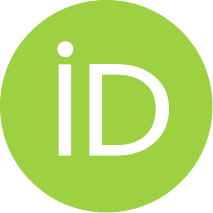}\hspace{1mm}Dongyang Kuang}\thanks{https://github.com/dykuang} \\
    School of Mathematics (Zhuhai)\\
	Sun Yat-sen University\\
	Guangdong, 519082, CHINA \\
	\texttt{kuangdy@mail.sysu.edu.cn} \\
	\And
    Xinyue Song \\
    School of Mathematics (Zhuhai)\\
	Sun Yat-sen University\\
	Guangdong, 519082, CHINA \\
	\texttt{songxy39@mail2.sysu.edu.cn} \\
	\And
	\href{https://orcid.org/0000-0002-6356-233X}{\includegraphics[scale=0.06]{orcid.pdf}\hspace{1mm}Craig Michoski} \\
	the Oden Institute for Computational Engineering and Sciences\\
	University of Texas at Austin\\
	201 E. 24th Street, POB 4.102 \\
    Austin, TX 78712, USA \\
	\texttt{michoski@oden.utexas.edu} \\
}

\hypersetup{
pdftitle={Emotion Classification from Multi-Channel EEG Signals Using HiSTN: A Hierarchical Graph-based Spatial-Temporal Approach},
pdfauthor={Dongyang Kuang, Xinyue Song and Craig Michoski},
pdfkeywords={Affective Computing, Emotion Recognition, EEG, Hierarchical Spatial Temporal Network, Parameter Efficient Models},
}

\begin{document}
\maketitle

\begin{abstract}
    This study introduces a parameter-efficient Hierarchical Spatial Temporal Network (HiSTN) specifically designed for the task of emotion classification using multi-channel electroencephalogram data. The network incorporates a graph hierarchy constructed from bottom-up at various abstraction levels, offering the dual advantages of enhanced task-relevant deep feature extraction and a lightweight design. The model's effectiveness is further amplified when used in conjunction with a proposed unique label smoothing method. Comprehensive benchmark experiments reveal that this combined approach yields high, balanced performance in terms of both quantitative and qualitative predictions. HiSTN, which has approximately 1,000 parameters, achieves mean F1 scores of 96.82\%  (valence) and 95.62\% (arousal) in subject-dependent tests on the rarely-utilized 5-classification task problem from the DREAMER dataset. In the subject-independent settings, the same model yields mean F1 scores of 78.34\% for valence and 81.59\% for arousal. The adoption of the Sequential Top-2 Hit Rate (Seq2HR) metric highlights the significant enhancements in terms of the balance between model's quantitative and qualitative for predictions achieved through our approach when compared to training with regular one-hot labels. These improvements surpass 50\% in subject-dependent tasks and 30\% in subject-independent tasks. The study also includes relevant ablation studies and case explorations to further elucidate the workings of the proposed model and enhance its interpretability.
\end{abstract}


\section{Introduction}\label{Sec:intro}

Initially proposed in \cite{picard2000affective}, the field of affective computing has since evolved to play a significant role within artificial intelligence. Among the multitude of data sources leveraged to discern human psychological states, non-invasive electroencephalogram (EEG) stands out due to its various advantages. These include but are not limited to its portability, relatively high temporal resolution, and assured safety. Integrating EEG-based affective computing with an array of pattern recognition tools, particularly the rapidly evolving neural network methodologies in deep learning, exhibits substantial potential across a multitude of applications \cite{gong2021deep,wang2022systematic,li2022eeg}. The preliminary step in comprehending human emotions is the construction of models that admit quantifiable parametric relationships.

Fundamentally, two abductive categories of models are usually identified to exist for quantifying emotions; the first, termed {\it discrete quantification models}, are often cited as being scientifically grounded in the early work of Ekman \cite{ekman2009darwin} and Plutchik \cite{plutchik2003emotions}, though are rooted in the work of the ancient philosophers, such as Aristotle \cite{aristotle1984}, Seneca \cite{seneca2023}, and Epictetus \cite{epictetus2023}. These models envision the emotion space as `patches', each representing a basic state such as, e.g., anger, anticipation, fear, sadness, disgust, trust, surprise, joy, etc., as well as their various combinations and embodiments. The second category of models (which might be viewed as a refinement of the first allowing for partial inclusions and multidimensional amalgams), are the {\it dimensional quantification models}, which employ mutually orthogonal axes to construct distinct (or independent) emotional dimensions. For example, within the realm of affective computing, Russell's Valence-Arousal bipolar emotional quadrant system \cite{russell1979affective} has gained wide acceptance, where the Valence axis aids in gauging an individual's happiness or sadness, while the Arousal axis quantifies the level of excitement. 
These geometry rich relations described by different emotion-based models provide many challenges for prediction frameworks in regular classification problems where OneHot labels are used to seek a model whose predictive behavior/logic is closer to that of human beings. A particular aspect of this concern is discussed more in part B of Section \ref{Sec:idea}.

Within the realm of numerical frameworks, recent successes of large-scale deep learning models, such as ChatGPT, have generated considerable interest in both industry and academia. However, in contrast to models such as ChatGPT, where huge datasets are available to train on, in the field of EEG-based human emotion recognition, a significant barrier impeding the effectiveness of large models in applications is data limitation. Despite researchers contributing to open-source datasets like SEED\cite{zheng2015investigating}, DEAP\cite{koelstra2011deap}, DREAMER\cite{katsigiannis2017dreamer}, ASCERTAIN\cite{subramanian2016ascertain}, etc., the overall volume of available data remains \textbf{extremely limited}, introducing a challenging problem when attempting to develop robust predictive and analytical frameworks. Nevertheless, extensive work on deep learning models in EEG-based emotion recognition has been done involving nearly all mainstream types of neural networks, including CNNs\cite{lawhern2018eegnet,wu2022simultaneously}, RNNs/LSTMs\cite{tao2020eeg,cui2020eeg}, capsule networks\cite{li2022emotion,wei2023tc}, graph convolution networks (GCNs)\cite{song2021graph,priyasad2022affect,liu2022bi,li2022residual}, transformers\cite{wei2023tc}, etc. In addition to these network models, active areas of study in the domain of EEG signal analysis also include: attention module designs \cite{kuang2023gram}, model compressing techniques \cite{liu2022bi}, and domain transfer learning approaches \cite{quan2023eeg}. It is, however, also worth pointing out that prediction tasks for fine-grained labeling approaches (e.g. valence level 1,2,...,5) are significantly less studied/understand than those of the more standard binary classification tasks (e.g, high v.s. low valence) commonly seen in benchmark studies. 

In addition to the frequent pursuit of enhanced predictive accuracy across diverse tasks, relatively fewer recent studies have concentrated on the qualitative characteristics of learning-based, data-driven models within the specific context of EEG-based emotion recognition. However, along these lines, in the pioneering work of \cite{lawhern2018eegnet} a compact model is proposed for accommodating the limited-data concern. In this work the spatial-temporal nature of  the signals is considered, but it is addressed via a simple approach by convolutional operations with kernels of custom sizes, which is more of a technique arising from general practices previously found within the computer vision community (note: in that context, though, without involving or adapting to EEG priors). Additionally, works like \cite{wu2022simultaneously}  further advance on the special spatial-temporal nature of EEGs by introducing a Multi-Scales Bi-hemispheric Asymmetric Model (MSBAM) that recasts the original EEG signal format into a sparse matrix representation of 3D input features by exploring a network design that incorporates the brain's bi-hemisphere asymmetry. The resulting MSBAM design, however, leads to a much larger network and the information learned from the ``empty'' entries in the constructed sparse input is still reliant on a significantly large latent space (i.e. conventional ``black-boxes). Comparatively, an example in work focused on using graph convolutional networks utilizes the technique of DiffPool \cite{ying2018hierarchical}, allowing for automatic graph hierarchies---though this approach was not originally intended for use on time series inputs. Because of this, notable augmentations and adaptations must be adopted to incorporate emotion-based EEG priors for enhancing the model performance. It is, however, also worth mentioning that none of these previous studies consider the question of whether the learned model's predictions (e.g. feature representation space) are inherently consistent between the different labeling models (i.e. for example, the aforementioned discrete and multidimensional emotion labeling models).

Drawing inspiration from prior research and identifying opportunities for enhancement within their methodologies, we introduce a Hierarchical Spatial Temporal Network (HiSTN) design. This approach seeks to harmonize the objectives of a lightweight model architecture with the establishment of a temporal-spatial hierarchy that integrates interpretable priors, while also ensuring that the model's predictive behavior aligns more closely with established clinical models of emotion. The key contributions of this paper are delineated as follows:

\begin{enumerate}
\item We propose a lightweight, parameter-efficient design tailored for prediction tasks that have limited training data.
\item We incorporate a hierarchical graph convolution component to extract spatial-temporal features at varying levels of abstraction. This design facilitates an intuitive interface for integrating prior knowledge about potentially useful hierarchical spatial information.
\item We introduce a special label smoothing technique that enhances the model's qualitative behavior, particularly in terms of continuity among the model's highest-ranked predictions. This technique helps to `shape' the learned feature representation space for tasks with categorical labels from multidimensional emotion models.
\end{enumerate}


The remainder of this paper is organized as follows: Section \ref{Sec:idea} provides a comprehensive explanation of our ideas for the network and training design, including the motivations behind these decisions from related work and a special label encoding method proposed for better consistency between numerical categorical labels and clinical emotion models during learning. Section \ref{Sec:exp} presents our main benchmark results for both subject-dependent and -independent tasks. This section also includes relevant case studies and ablation studies that investigate the effects of some crucial choices made during model specification and training strategies.

\section{The Motivation and Idea}\label{Sec:idea}
\subsection{Network Design}
Generally speaking, EEG signals recorded from non-invasive devices have relatively good temporal resolution. The regular time convolution layers, such as a one dimensional convolution layer, tend to be quite effective for extracting time related features hierarchically when being properly stacked in a network design framework. On the other hand, spatial information is often considered lacking in non-invasive EEG signals. Due to the limitation in EEG signals, and the considerable utility and value of spatial information, much research has been performed aimed at enhancing the relatively poor spatial resolution of EEG signals, and then designing these enhancements into the network construction process itself. 

For example, prior studies, exemplified by \cite{song2021graph,li2022spatial}, have utilized diverse projection and interpolation techniques to convert multi-channel time series data into a format compatible with image-based data. In a similar vein, \cite{wu2022simultaneously} introduced a method that arranges distinct channels into a sparse 2D matrix representation, in order to reveal certain spatial signal components. However, the emergence and growing acceptance of graph convolutional neural networks (GCNs) have prompted researchers to adopt this framework with increasing regularity when developing spatially resolving models.

There are at least \textit{two crucial issues} that arise from the GCN approach. The first pertains to \textit{the construction of an appropriate graph}. Previous practices encompassed the use of manually specified graphs based on clinical priors \cite{wang2019convolutional,tian2022applying}, employing different functional connectivity measures, or dynamically generating/adjusting them during the learning process \cite{song2018eeg,song2021graph,priyasad2022affect}. The second key concern arises during the graph embedding step. Traditionally, this step involves Multi-Linear Perceptron (MLP) layers that map node features at layer $l-1$ with dimension $d_{l-1}$ to the desired node feature dimension $d_{l}$ at layer $l$ (where $d_{l-1}$ and $d_{l}$ may be equal) in the subsequent layer. However, if one tries to utilized GCN with signals directly as node features, as determined by the sampling frequency and time window, the GCN layer would accumulate \textit{a significant number of parameters} during this embedding process. To mitigate this, related work either conducts manual feature extractions beforehand and employs them as input instead of the original signal \cite{priyasad2022affect,li2022spatial}, or positions the graph layer in deeper blocks of the network design, where the time dimension of learned abstract features is reduced by previous time convolutional blocks \cite{song2021graph}. Our work in this paper aligns with these ideas, but with more focuses on a light-weighted yet structure-rich graph design.  

\begin{figure}[tbh]
    \centering
    \includegraphics[width=.75\textwidth, height=230pt]{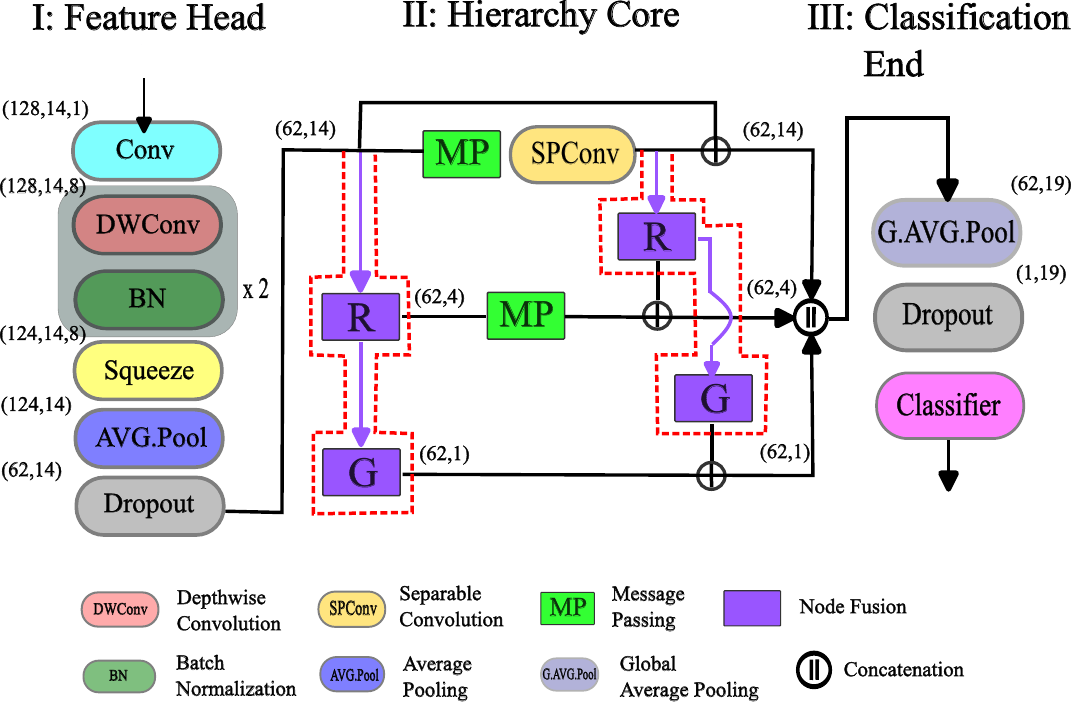}
    \caption{The design of the proposed HiSTN network. A closer look at the Hierarchy Core (enclosed by the red dashed line) is unpacked in Fig. \ref{fig:node_fusion}. }
    \label{fig:HiSTN}
\end{figure}

\begin{figure}[tbh]
    \centering
    \includegraphics[width=0.15\textwidth]{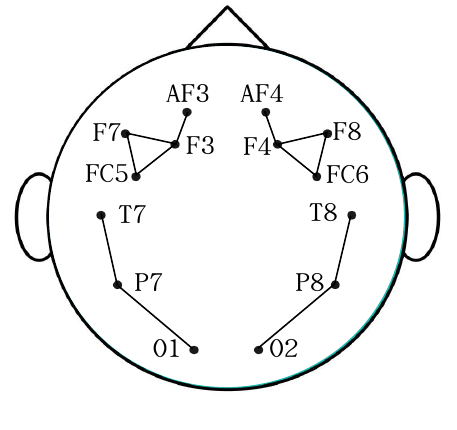}
    \includegraphics[width=0.3\textwidth]{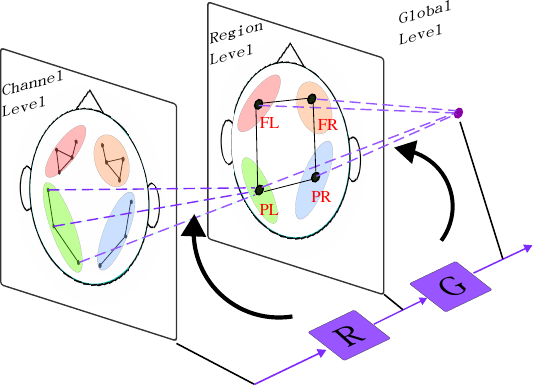}
    \caption{Unpacking the node fusion block. FL: Frontal Left, FR: Frontal Right, PL: Parietal Left, PR: Parietal Right. At the intermediate stage, Region block R ``summarizes" the learned information from previous node/channel level features. This processed information per region is then further summarized by the Global block G over the whole graph. }
    \label{fig:node_fusion}
\end{figure}


Our overall design, as depicted in Fig. \ref{fig:HiSTN}, consists of three main parts: a \textbf{`Feature Head'} for successive feature preprocessing and auto-extraction (I), a \textbf{`Hierarchy Core'} for developing information hierarchy via proper graphs (II) and a \textbf{`Classification End'} (III). In line with previous research like EEGNet \cite{lawhern2018eegnet}, the `Feature Head' comprises a series of stacked temporal convolution layers for extracting lower-level features primarily focused on temporal aspects. This is achieved by using convolution kernels of shape $(k,1), k > 1$ (in all convolution layers) limiting the convolution to the time dimension. It is also worth noting that the very first 2D convolution layer expands a typical spatial-temporal signal of shape $(T,C)$ to $(T,C,S)$ by learning $S$ different convolution kernels that provide the ``multiple view:'' $X^m \in \mathcal{R}^{T\times C \times S}$ for further feature extraction. That is, the squeeze layer learns proper weights for summarizing previous multiple views into one via setting $X^{squeezed} = \sum_{i=1}^S w_iX^m_i$ where $X^m_i$ denotes the $i$th view. This also facilitates dimension reduction along the time axis via temporal pooling, resulting in reduced computational complexity during the graph message passing stage in Part II. The `Hierarchical Core' enriches the hierarchical spatial information by establishing different levels of graphs (channel level, region level, and global level) through node fusion blocks. Message passing is performed on the corresponding graphs at each level. The nodes at the channel level correspond to individual EEG channels, while the nodes at the region level and graph level are abstract concept nodes. The information flow within this block will be introduced below. Finally, Part III is a standard `Classification End' consisting of pooling, dropout layers, and a relevant classifier.

Fig. \ref{fig:node_fusion} offers insight into the pathway (highlighted in purple and enclosed by the red dashed line in Fig. \ref{fig:HiSTN}) responsible for constructing a graph hierarchy. An example is presented by partitioning all channels into four regions. The graph formed by AF3-F3-F7-FC5 is denoted as FL, representing the Frontal Left region. Similarly, PL represents T7-P7-O1, encompassing the temporal/parietal/occipital left region. FR and PR are defined similarly for regions on the right hemisphere. This hierarchical graph structure is illustrated in Fig. \ref{fig:HiSTN}, with tensor shapes marked to aid readers in tracking the flow of information. For instance, the node fusion block labeled with the letter ``R" is responsible for fusing channel-level feature of shape (62, 14) into region-level features of shape (62, 4). The fusion block, denoted as ``G", further abstracts region-level features to form graph-level features of shape (62, 1). These hierarchical deep features learned at different spatial levels are then concatenated resulting in a feature tensor of shape (62, 19), which is then globally pooled along time dimension forming a summary feature vector of length 19 before passing to the classifier in Part III.

Information flow occurs between different nodes based on the underlying graph structure at each level, with information being transmitted or fused exclusively from lower levels to higher levels to emulate the process of abstraction. While a fully connected graph can be used at the channel level, we opt for a graph comprising $N$ connected components for a couple reasons. First, it prevents potential interference from nodes belonging to other regions during feature fusion at the next level. Second, it simplifies the graph structure and facilitates future parallel implementation. This design offers a convenient interface for incorporating specific prior knowledge about functional connectivity graphs. Further exploration of graph choices is conducted in Section \ref{sec:ablation} through ablation studies. 

The feature fusion from the previous to the subsequent layer (purple dashed line in Fig. \ref{fig:node_fusion}) can be described as follows:
\begin{eqnarray}
   X^{(l)}_i  = \sum\limits_{j \in N^{(l-1)}_i } C^i_j X^{(l-1)}_{j}, \ \ C^i  = \varphi\left( {\rm MLP} ( X^{(l-1)}_{i} ) \right),
\end{eqnarray}
such that $X^{(l)}_i$ is the node $i$'s feature at level $l$ formed by the weighted sum from its corresponding node features $X^{(l-1)}_{j}$ at level $l-1$. The collection of node $i$'s neighboring nodes at the previous level $l-1$ is denoted by $N^{(l-1)}_i$. The learnable weight vector $C^i = [C^i_1, \cdots, C^i_{N_i}]$ correspond to the outputs from a single MLP layer activated by the regular softmax function $\varphi$. As for the graph message passing within each graph convolution level, we adopt the following method using a Chebyshev polynomial on the normalized graph Laplacian matrix as the graph convolution kernel for faster and more stable approximations \cite{defferrard2016convolutional,he2022convolutional}: 
\begin{eqnarray}
    & \hat{X}^{(l)}_i = \sigma \left( \sum\limits_{k=0}^d \beta_k T_k(\tilde{L}) X^{(l)}_i \right),\\ 
    & \mathrm{where}\  L = D - A,  \ \tilde{L} = \frac{2L}{\lambda_{\max}} - I.
\end{eqnarray}
Here, $\sigma$ is a nonlinear activation function, $A$ is the graph adjacency matrix, $D$ is the corresponding diagonal matrix with diagonal entries being the degree of the corresponding node, and $\lambda_{max}$ is the maximum eigenvalue of $L$. The $k$-th order Chebyshev polynomial can be obtained conveniently via the usual recurrence formula:
\begin{eqnarray}
    & T_k(\tilde{L}) = 2\tilde{L}T_{k-1}(\tilde{L}) - T_{k-2}(\tilde{L}), \ 
    \\ & \mathrm{where} \ T_0(\tilde{L}) = I \ \mathrm{and} \ T_1(\tilde{L}) = \tilde{L}. \ 
\end{eqnarray}

In our experiments in Section \ref{Sec:exp}, the maximum degree $d$ is chosen to be the diameter for each graph considered. As for the feature processing after message passing, we use a 1D separable convolution along the feature's time dimension instead of dense layers for capturing possible time related information in the nodal features. With the adopted graph shown in Fig. \ref{fig:HiSTN}, 
features learned from these three levels are later concatenated together. The concatenated feature is then globally averaged/pooled along the time dimension before being fed to the classifier. 
It is important to highlight key distinctions between our approach and other works employing graph hierarchies, such as DiffPool \cite{ying2018hierarchical}, which focuses on general and automatic graph pooling for hierarchical learning. DiffPool initiates with a connected graph and dynamically learns assignment matrices for node fusion, adjacency matrices for message passing, and feature embedding matrices during training. Each node at level $l-1$ can be associated with all nodes at the next level, which differs from our aforementioned design. Additionally, instead of regular graph embedding via matrix multiplication, we utilize a separable convolution along the time dimension for feature embedding, considering that the features at each node are time series features. Time convolution ($\mathcal{T}$)\footnote{This time convolution is performed as depthwise convolution in implementation.} requires fewer parameters and is generally faster than MLP. However, it does not commute with the message passing operation ($\mathcal{\psi}$) mathematically in general, i.e. $\mathcal{T}\circ \mathcal{\psi} \not\equiv \mathcal{\psi}\circ \mathcal{T}$
, prompting us to adopt the multi-branched design in Part II of Fig. \ref{fig:HiSTN} for feature fusion before and after the aforementioned operations. 


\subsection{Prediction Framework}\label{sec:prediction framework}
One of the goals in current and future deep learning tools is to make prediction models more human-like, or more capable of emulating clinical human responses or researcher-compatible `interpretive evaluations.' For example, in the context of training models for EEG-based emotion recognition, especially for score-based predictions, given a particular dataset/context with a fixed emotional stimulus `type', 
there often exists certain levels of uncertainty/fuzziness in the self-assessed rating. For instance, if someone's self-assessment during a emotional stimuli type is 6 on a 1-9 scale, the `subjectivity' of personal perception, together with the natural order of the scale, implies nearby scores of 5 or 7 are more likely when tested at a different time than more extreme variations such as a score of a 1 or a 9. A standard OneHot labeling approach does not address these logical priors well and it can lead to an over-confidence concern when training a model \cite{pmlr-v70-guo17a}. As an easy and straightforward approach for incorporating this logical prior into the model, which we refer to as the ``Continuum of Predictions (CoP)" behavior for trained models, a prior distribution centered at the maximum likelihood of the existing self-assessed scores can be utilized for modifying training labels. While previous prediction methods have primarily focused on top-1 accuracy or similar performance metrics, the consideration of this ``Continuum of Predictions" has received less attention, especially under fine-grained predictions which we adopted for model comparisons in this paper. It is worth noting here that while loss functions such as mean squared error or mean absolute error, instead of the widely adopted cross-entropy loss for classification tasks, can naturally address the prediction continuum problem, research in the field of ``learning with noisy labels" has shown that these loss functions' generalization performance significantly degrades when dealing with complex data, compared to categorical cross-entropy (CCE) loss \cite{ghosh2017robust,song2022learning}. Thus, as a second objective in this paper, we present an easy variation of the classical label smoothing technique \cite{szegedy2016rethinking,muller2019does} to address the issue of CoP under the task of fined-grained emotion score predictions. To enable better comparison, we explore four different classifier designs in this work.
\begin{itemize}
    \item[\textbf{A}:] The model is trained using regular OneHot label encoding, where the label $2$ is encoded as $\{0,0,1,0,0\}$, and trained with Categorical Cross-Entropy (CCE) loss between true labels $y$ and the associated predictions $\hat{y}$: $L(y,\hat{y}) = -\sum\limits_{i=1}^N y_i \log(\hat{y}_i)$.
    \item[\textbf{B}:] The model's output directly predicts the subject's self-reported score, and it is trained using mean absolute error (MAE) loss: $L(y,\hat{y}) = -\sum\limits_{i=1}^N | y_i - \hat{y}_i |$. 
    \item[\textbf{C}:] The model's output consists of the parameters $\theta$ describing a Gaussian Mixture Model (GMM) with 5 components ($N=5$), and it is trained using Negative Log-likelihood Loss, $L(y,\hat{y}) = - \log(p_{\mathrm{GMM}}(\hat{y}; y,\theta))$, based on the GMM probabilities.
    \item[\textbf{D}:] The model is trained using a specially smoothed label encoding method (see Eq-\ref{eq:smooth} below). For example, \[\{ 0,0,1,0,0\} \rightarrow \{2.64\times10^{-4}, 0.11, 0.79, 0.11, 2.64\times10^{-4}\}\] represents the smoothed encoded label for score 2. The model is also trained with Categorical Cross-Entropy (CCE) loss. This is the proposed label encoding method.
\end{itemize}

Eq-\ref{eq:smooth} gives the formula smoothing the label for addressing the prediction `continuum' problem, where $i$ is the true label and $j$ is the index corresponding to the $j$-th score in the smoothed label. The $P_j$ gives the smoothed value\footnote{It can also be viewed as a fuzzy membership value on the label set.} at index $j$, while $s=0.5$ (half the width between two consecutive ratings) is used in our study here, yielding:
\begin{equation}\label{eq:smooth}
    P_j =  \frac{\exp(-(j-i)^2/2s^2)}{\sum_{j} \exp(-(j-i)^2/2s^2)}.  
\end{equation}
Our benchmark experiments in Section \ref{Sec:exp} have shown that this simple modification of label encoding can greatly help improve the resulting model's top predictions. Codes relevant with this paper will be made available on Github at \url{https://github.com/dykuang/EEG-based-affective-computing}.

\section{Experiment}\label{Sec:exp}
\subsection{The DREAMER Dataset}

The DREAMER Dataset \cite{katsigiannis2017dreamer} is a multimodal database containing EEG and ECG signals recorded during the elicitation of affect using audio-visual stimuli. The dataset comprises data from 23 subjects (14 males and 9 females), including their self-assessments (integers from 1 to 5) in terms of valence, arousal, and dominance after each stimulus. The dataset consists of 18 film clips, with varying durations (ranging from 65 seconds to 393 seconds), and for detailed information about each film clip, readers can refer to \cite{gabert2015ratings}. It is worth noting that some subjects' labels do not cover the full range of scores for arousal and dominance, but only a subset of them. For the EEG signal collection, the Emotive EPOC wireless headset and the Shimmer 2 ECG sensors were utilized. 

In this particular experiment, we solely utilized the EEG signals, which consists of a total of 14 channels, namely AF3, F7, F3, FC5, T7, P7, O1, O2, P8, T8, FC6, F4, F8, and AF4. 
Previous research studies have demonstrated promising results in binary classification tasks using this dataset, considering both the subject-dependent and the subject-independent settings (refer to Table \ref{tab:Dreamer-prev}). For binary classifications, a threshold (such as 3) is selected to map the original 5 scores into 2 classes (high vs. low). The works presented in Table \ref{tab:Dreamer-prev} can differ in some details about experimental settings, including training data preparation, normalization methods, and evaluation criteria, among others.  

\begin{table}[tbh]
    \centering{
        \caption{A collection of previous and recent work on the binary classification task using DREAMER. The Percentage refers to accuracy (mean $\pm$ std.dev.). Rows corresponding to subject-independent experiments are colored in cyan. The Notes column collects the evaluation method and length of signals used for prediction.}
        \label{tab:Dreamer-prev}

     \begin{tabular}{|c|c|c|c|c|}
        \hline
        Models & Year & Valence(\%) & Arousal(\%) & Notes\\
        \hline
        GECNN\cite{song2021graph} & 2021 & 95.73 $\pm$ - & 92.79 $\pm$ - & leave one trial out CV, 2s\\
        \hline
        DCNN+GAT-MHA\cite{priyasad2022affect} & 2022  & 88.80 $\pm$ -& 88.24 $\pm$ -  & 10CV, 1s\\
        \hline
        SFCSAN\cite{li2022spatial} & 2022 & 93.77 $\pm$ - & 95.80 $\pm$ -  & 10CV, 1s\\
        \hline
        MTCA-CapsNet\cite{li2022emotion}& 2022 & 94.96 $\pm$ 3.60 & 95.54 $\pm$ 3.63  & 10CV, 1s\\
        \hline
        ACRNN\cite{tao2020eeg} & 2022 & 97.93 $\pm$ 1.73 & 97.98 $\pm$ 1.92  & 10CV, 1s\\
        \hline
        RGCB\cite{li2022residual} & 2022 & 87.43 $\pm$ 14.89 & 91.55 $\pm$ 14.78  & leave one session out CV, 1s\\
        \hline
        Bi-CapsNet\cite{liu2022bi} & 2023 & 95.48 $\pm$ 3.26 & 95.86 $\pm$ 3.31  &  10CV, 1s\\ 
        \hline
        TC-Net\cite{wei2023tc} & 2023 & 98.59 $\pm$ 1.38 & 98.61 $\pm$ 1.34   & 10CV, 1s\\
        \hline
        MSBAM\cite{wu2022simultaneously} & 2023 & 99.69 $\pm$ 0.24 & 99.76 $\pm$ 0.20   & 10CV, 1s\\  
        \hline
        TDMNN \cite{ju2023eeg} & 2023 & 99.45 $\pm$ 0.51 & 99.51 $\pm$ 0.79   & 5CV, 3s \\
        \hline
        \hline
        \rowcolor{LightCyan} RMCNN \cite{maheshwari2021automated} & 2021 & 58.02 $\pm$ - & 51.23 $\pm$ -  & LOOCV, 10s\\
        \hline
        \rowcolor{LightCyan} SparseD \cite{zhang2021sparsedgcnn} & 2021 & 64.06 $\pm$ 8.58 & 66.96 $\pm$ 6.91  & LOOCV, 2s\\
        \hline
        \rowcolor{LightCyan} FLDNet \cite{wang2021fldnet} & 2021 & 89.91 $\pm$ 12.51 & 87.67 $\pm$ 10.02  &  5CV, 3s\\
        \hline
        \rowcolor{LightCyan} DCNN+GAT-MHA\cite{priyasad2022affect} & 2022  & 64.98 $\pm$ - & 63.71 $\pm$ -  & LOOCV, 1s\\
        \hline
        \rowcolor{LightCyan}  MR-VAE-DT \cite{quan2023eeg} &  2023 & 73.08 $\pm$ 8.84 & 72.74 $\pm$ 12.93 & LOOCV, 1s\\			
        \hline
        \rowcolor{LightCyan} DCNN+NN \cite{singh2023deep} & 2023 & 96.74 $\pm$ - & 97.64 $\pm$ -  & 80/10/10 split, 1s\\
        \hline       
    \end{tabular}
    }       

\end{table}

This paper focuses on the more challenging and less studied task of \textbf{5}-classification for better assessing not only the proposed network's learning ability, but also the qualitative improvement our proposed label smoothing trick can bring to the prediction. To account for the time it can take for emotions to develop \cite{lerner2015emotion,adolphs2002neural}, not all data is utilized in our benchmark evaluations. Following previous research as in Table \ref{tab:Dreamer-prev}, we consider the last 1 minute of data during the stimuli for our experiments. The data is normalized using the corresponding 1-minute baseline signal when no stimuli are presented. In our benchmarks, we also include EEGNet as a representative compact model baseline and MSBAM as an example of a larger network utilizing matrix embedding and bi-hemisphere asymmetrical priors to enhance spatial information for comparison.

The F1 score is utilized to evaluate the top prediction performance across different configurations. Additionally, the top-2 accuracy is employed to further compare the predictive capabilities of trained models. In addition to commonly used quantitative measures, we introduce two additional metrics for assessing the qualitative behavior of the models in terms of the mentioned ``Continuum of Prediction". The first metric, referred to as ``Tridiagonal Percentage (Tri-P)", is calculated as 
$\text{Tri-P} = 100\times\left(\frac{\sum_{|i-j|<2} C_{ij}}{\sum_{i,j} C_{ij}}\right)$ where $C$ represents the confusion matrix,
and this quantity represents a percentage ratio between the sum of elements that are on the diagonal, subdiagonal, or superdiagonal and the number of total test samples. The higher the metric is, the better the overall prediction quality is in terms of CoP. The second metric measures the proportion of samples in the test set that meet two conditions: (1) the model's top-2 predictions for the sample are consecutive, and (2) the true label for the sample falls within the model's top-2 predictions. In this paper, we refer to this metric as the ``Sequential Top-2 Hit Rate" or `Seq2HR'. A higher value indicates better overall prediction performance. Within each benchmarked model, we will compare the four variations with different classifier designs as mentioned in Section \ref{Sec:idea}.
All numerical experiments are performed with \textit{Tensorflow} framework on a \textit{Nvidia 3080Ti} graphic card.  

\subsection{Subject-Dependent Experiments}\label{Sec:exp-sd}
\begin{wraptable}{l}{0.25\textwidth}
    \caption{Hyper-parameters used in benchmark.}
    \label{tab:HP-SD}
    \begin{tabular}{lc}
    \hline
        Batchsize &  256\\
        LearningRate & $0.001$\\
        Maximum Epochs & 50\\
    \hline
    \end{tabular}
\end{wraptable}
As mentioned earlier, even if some subjects' self-reported labels contain only a subset of the 5 ratings, we maintain consistency in predictions among all subjects by using the same model configuration with 5 output classes. A special 10-fold cross-validation (10 CV) approach is employed for the benchmarks in this section. The 1-minute data is divided into 10 consecutive non-overlapping trunks, each lasting 6 seconds. During each iteration, one fold is used for testing, the fold preceding it is used for validation, and the remaining folds are used for training. To address data limitations and potential label imbalance issues, a data generator is utilized to randomly select 1-second segments from the training data. Within each batch, an equal number of samples is associated with each different label. The model with the highest validation performance is saved, and its performance is evaluated on the test set, which consists of 1000 randomly generated 1-second samples from the mentioned data generator. The Adam optimizer is utilized for training, and additional important hyperparameter settings are detailed in Table \ref{tab:HP-SD}. The benchmark performance for the 5-class prediction, under different model configurations, is summarized in Table \ref{tab:Sd}. The best performance per metric column overall is highlighted in \textbf{bold}, while the best performance for each model among different training variations is colored with an orange background. The second-best performance for each model is colored with a cyan background.

\begin{table}[tbh]
    \centering
    \caption{Summary of subject-dependent experiments on 5-classification tasks. }
    \label{tab:Sd}
    \begin{adjustbox}{width=\columnwidth,center}
    \begin{tabular}{|p{5mm}|c|c|c|c|c|c|c|c|c|c|}
    \hline
    \multicolumn{3}{|c}{ } &\multicolumn{4}{|c}{Valence}  & \multicolumn{4}{|c}{Arousal} \\
    \hline
        \multicolumn{2}{|c|}{Models} & \#Para.  & F1(\%)  & Top2 Acc.(\%) & Tri-P(\%) & Seq2HR (\%)& F1(\%)  & Top2 Acc.(\%) & Tri-P(\%) & Seq2HR (\%)\\
         \hline
        \multirow{3}{3em}{\tiny{EEGNet}} 
                                  &  A &  3269 & \myccLC 94.74 $\pm$ 3.82 &  \myccO 98.98 $\pm$ 0.81&  \myccLC 97.09 $\pm$ 2.35 & 42.80 $\pm$ 13.54 & \myccLC 95.59 $\pm$  3.40  & \myccLC 99.01 $\pm$ 1.06 &  \myccLC 98.28 $\pm$ 1.71 & 52.19 $\pm$ 18.07\\ 
                                  &  B &  1217 &63.22 $\pm$ 14.31 & 86.10 $\pm$ 9.37 & 93.57 $\pm$ 4.76 &\myccLC 86.10 $\pm$ 9.36 &67.29 $\pm$ 17.24 & 89.53 $\pm$ 9.30 & 95.48 $\pm$ 5.44 & \myccLC 88.63 $\pm$ 10.22\\
                                  &  C &  8399 & \myccO 95.10 $\pm$ 4.16 &  \myccLC 98.78 $\pm$ 1.33& \myccO 98.00 $\pm$ 1.88 & 59.48 $\pm$ 15.39 & \myccO 96.02 $\pm$ 3.22&   \myccO 99.18 $\pm$ 0.64 & \myccO 98.54 $\pm$ 1.43 & 62.86 $\pm$ 18.11\\
                                  &  D &  3269 &91.13 $\pm$  6.00&  92.44 $\pm$  2.41&  96.69 $\pm$ 3.16 & \myccO 91.71 $\pm$ 6.30 &90.95 $\pm$ 5.80  &  97.76 $\pm$ 1.81 &  97.07 $\pm$ 2.36 & \myccO 92.72 $\pm$ 5.58\\
        \hline
        \multirow{3}{3em}{\tiny{MSBAM}} 
                                  &  A & 203,269 &92.70 $\pm$ 3.96  &  97.70 $\pm$ 1.61&  95.73$\pm$ 2.55 & 41.46 $\pm$ 13.94 &92.90 $\pm$ 4.76  & 98.06 $\pm$  1.85 & 97.28 $\pm$ 2.47 & 54.36 $\pm$ 19.35 \\ 
                                  &  B &  202,785 &88.49 $\pm$ 6.06   & 96.15 $\pm$ 2.51 & \myccLC 98.07$\pm$ 1.34 & \myccO 96.15 $\pm$ 2.51 &89.60 $\pm$ 7.44  & \myccLC 96.95 $\pm$ 3.02 & \myccLC 98.62 $\pm$ 1.61 & 96.93 $\pm$ 3.02 \\
                                  &  C & 204,479 & \myccO 95.77 $\pm$ 2.49 & \myccO 98.29 $\pm$ 1.23 & 98.03$\pm$ 1.36 & \myccLC 72.10 $\pm$ 10.53 & \myccLC 95.26 $\pm$ 3.03  & \myccLC 98.30 $\pm$ 1.20 & 98.46 $\pm$ 1.37  & 64.42 $\pm$ 13.71 \\
                                  &  D & 203,269 & \myccLC 95.29 $\pm$ 3.01 & \myccLC 98.26 $\pm$ 1.28 & \myccO 98.16$\pm$ 1.53 & 96.11 $\pm$ 3.16 & \myccO 95.63 $\pm$  2.68  & \myccO 98.47$\pm$ 1.34 & \myccO 98.71 $\pm$ 1.29 & \myccO 97.51 $\pm$ 2.35\\
        \hline
        \multirow{3}{3em}{\tiny{HiSTN}}    
                                  &  A & 1181 & \myccO \textbf{97.13} $\pm$ \textbf{1.60}  &  \myccO \textbf{99.59} $\pm$ \textbf{0.29} &  \myccLC 98.49 $\pm$ 0.95 & 45.22 $\pm$ 12.88 & \myccO \textbf{97.33} $\pm$ \textbf{2.20} &  \myccO \textbf{99.67} $\pm$ \textbf{0.45} & \myccLC 98.97 $\pm$ 1.43 & 47.72 $\pm$ 19.46 \\ 
                                  &  B & 1101 &82.08 $\pm$ 6.48   &  95.98 $\pm$ 1.99 & 98.41 $\pm$ 0.98 & \myccLC 95.98 $\pm$ 1.98 &85.61 $\pm$ 8.74 & 96.73 $\pm$ 3.53 & 98.82 $\pm$ 1.59 & \myccLC 96.70 $\pm$ 3.53 \\
                                  &  C & 1381 &94.13 $\pm$ 3.16  &  98.77 $\pm$ 1.00 &  98.42 $\pm$ 0.97 & 75.80 $\pm$ 11.52 &94.53 $\pm$ 4.51 & 98.81 $\pm$ 1.22 & 98.49 $\pm$ 1.57 & 69.34 $\pm$  12.25 \\
                                  &  D & 1181 &\myccLC 96.82 $\pm$ 1.65  &  \myccLC 99.28 $\pm$ 0.53 &  \myccO \textbf{99.22} $\pm$ \textbf{0.72} & \myccO \textbf{97.57} $\pm$ \textbf{2.15} &\myccLC95.62 $\pm$ 2.96 &  \myccLC 99.23 $\pm$ 0.67 &  \myccO \textbf{99.29} $\pm$ \textbf{0.80} & \myccO \textbf{97.82} $\pm$ \textbf{1.96}\\
         \hline
    \end{tabular}
\end{adjustbox}
\end{table}

Among the three compared network designs, MSBAM has the highest number of parameters (approximately 200k) among the selected architectures, but it is still not considered large compared to most other methods listed in Table \ref{tab:Dreamer-prev}. While EEGNet is already designed to be compact, the proposed HiSTN used in this experiment significantly reduces the parameters to only about \textbf{\textit{1k}}. In terms of F1 score, HiSTN-A (HiSTN with regular OneHot encoding) achieves the highest values for both valence (\textbf{97.13\%}) and arousal (\textbf{97.33\%}) prediction. HiSTN-D (HiSTN with specially smoothed label encoding) outperforms other configurations in terms of other metrics, particularly showing notable improvements (at least \textbf{50\%}) for Seq2HR. Similar improvements are observed with EEGNet and MSBAM, indicating the universal effectiveness of enhancing the prediction's ``continuum'' behavior\footnote{Pairwise t-test gives very small p-values (i.e. EEGNet: $4.92e^{-14}/4.09e^{-11}$, MSBAM:$4.28e^{-14}/2.03e^{-11}$, HiSTN: $2.25e^{-14}/1.59e^{-11}$, for valence/arousal respectively, suggesting strong statistical significance) when testing if the proposed special label smoothing can bring improvements to Seq2HR against regular OneHot label encoding. }.

Although conducting classification as a regression task (training type B) ensures perfect prediction ``continuum" behavior, the distance-based loss is not as effective as the softmax-typed loss in training the model for accuracy. For all three models considered, the corresponding F1 scores and Top2 accuracy are the lowest among the four training variations compared. This suboptimal accuracy behavior also impacts other metrics such as Tri-P and Seq2HR. When comparing MSBAM and HiSTN with EEGNet, notable parameter differences can be observed between training variations B and C, primarily influenced by the transition from flattened features to the last dense layer. The number of parameters is sensitive to the hidden units in the last layer, with EEGNet exhibiting this sensitivity more prominently. MSBAM and HiSTN, on the other hand, demonstrate a more balanced distribution of parameters across shallow and deep layers, resulting in reduced sensitivity to the hidden units in the last dense layer. While training variation C (model outputting a mixed Gaussian distribution) shows improved Seq2HR despite having the most parameters, it remains competitive across other metrics. In the experiments with EEGNet, it achieves the highest F1 scores and Top2 accuracy for both valence and arousal score prediction tasks. For experiments with MSBAM, it achieves the highest F1 scores and Top2 accuracy in the valence score prediction task and the second highest (very close to the highest value) in the arousal score prediction task.

\begin{figure}[tbh]
    \centering
    \includegraphics[width=0.45\textwidth, height=300pt, trim=40 50 20 40,clip]{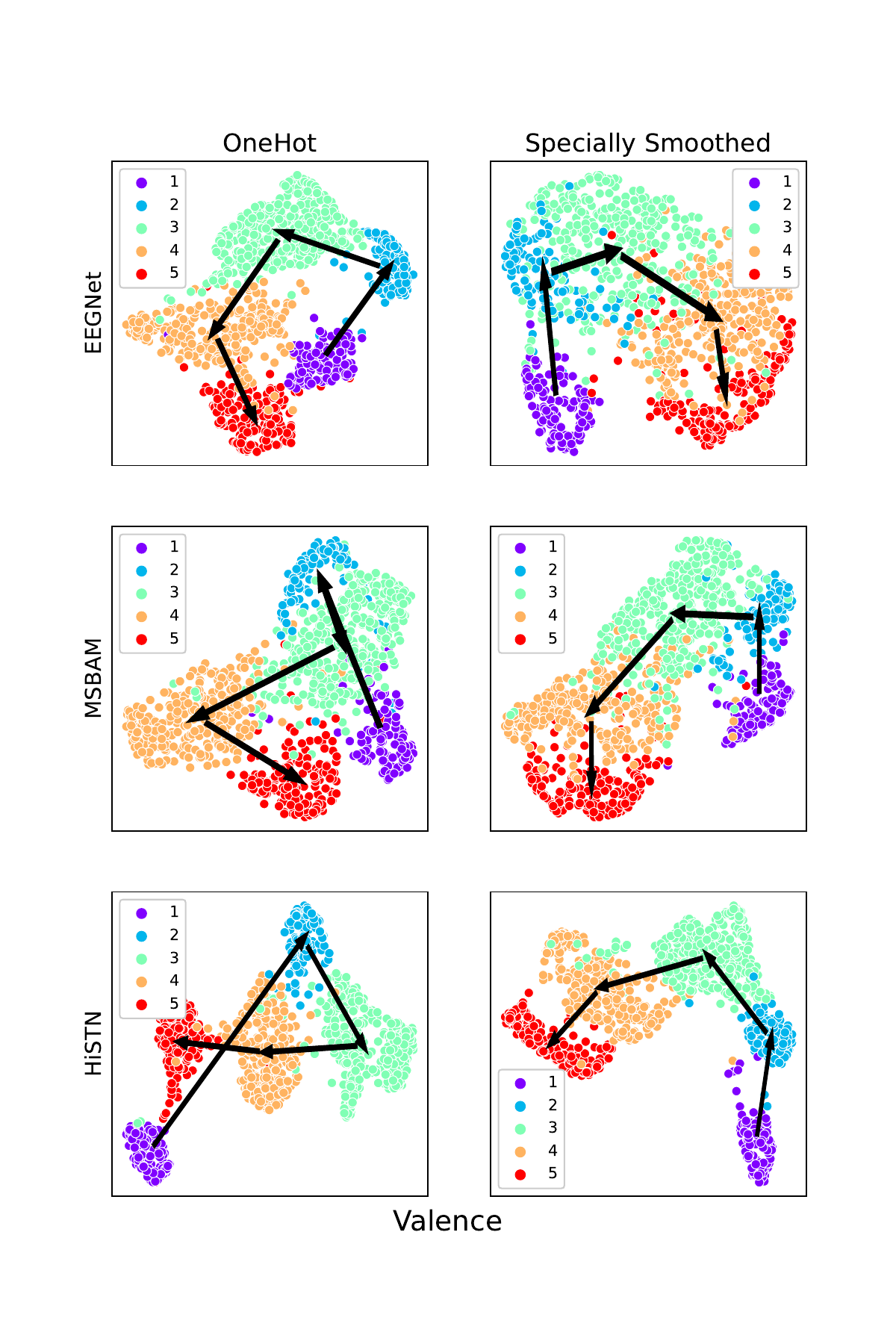}
    \vline height 270pt depth 0pt width 1pt
    \includegraphics[width=0.45\textwidth, height=300pt, trim=40 50 20 40,clip]{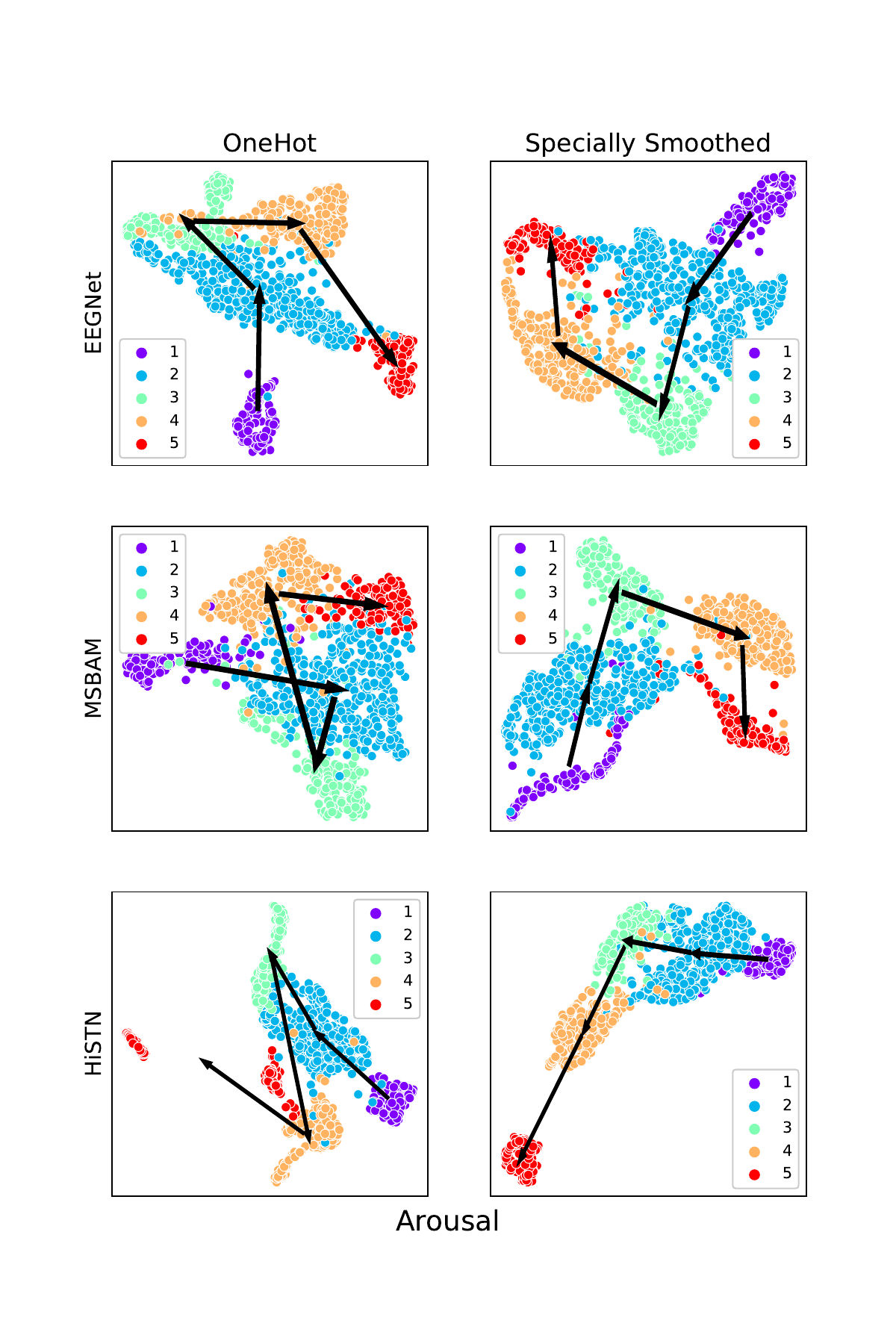}
    \caption{2D embedding of deep features using UMAP when different models are trained with (1) regular OneHot label encoding, and (2) our proposed special label-smoothing. Left: Valence prediction with date from Subject S3, Right: Valence prediction with data from Subject S23.}
    \label{fig:features}
\end{figure}
\subsubsection{Case Study -- Representation Space}
As discussed and validated in Table \ref{tab:Sd}, training with our proposed specially smoothed labels can significantly help increase the prediction quality in terms of CoP. To provide a different visualization perspective for gauging the improvements these smoothed labels can bring to the learned feature representation space, we also present Fig. \ref{fig:features}. In this figure, features output by the last dense layer before activation are embedded into two dimensions via the UMAP algorithm \cite{umap}.  In addition, arrows from the cluster center ranked $i$ to cluster centers ranked $i+1$ are also attached, as further shown in Fig. \ref{fig:features}.  From this plot, one can easily see that the manifold bearing representations belonging to different ranking scores ($1 \rightarrow 5$) is more intuitive---specially there are no self-intersections---after training with smoothed labels. Moreover, clusters ranked 1 and cluster ranked 5 are more separated compared with the cases utilizing regular OneHot labels. These observations help support the conclusion that the representations learned with the proposed smoothed labels are better able to model human intuition and logic on at least two aspects; i.e. \textbf{\textit{1)}} representations corresponding to rankings from 1 through 5 are properly aligned on the representation manifold with their natural 1-D order, and \textbf{\textit{2)}} clusters with rankings 1 and 5 are visually easier identified as the two ends on the representation manifold when compared to clusters with rankings 2 through 4, which comprise the interior points.

\begin{figure}[tbh]
    \centering
    \includegraphics[width=0.75\textwidth]{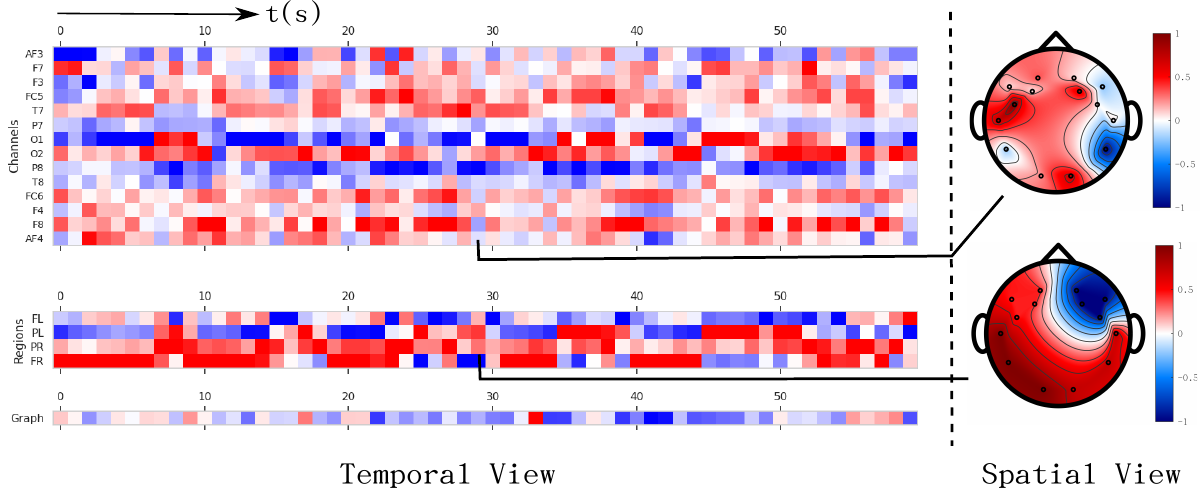}
    \caption{Deep features extracted in the temporal view and the spatial view at a time snapshot. From top to bottom: channel level, region Level and graph Level. The spatial view for the graph level is not shown since it is a scalar value.}
    \label{fig:TSfeature}
\end{figure}

\subsubsection{Case Study -- Deep Features}
Fig. \ref{fig:TSfeature} provides a visual example of learned deep features before the classifier layer (i.e. a vector of 19 dimensions: 14[No. channels]+4[No. regions]+1[No. graphs] per 1s input as seen in Fig. \ref{fig:HiSTN}), sequentially stacked along time representing the entire 1-minute recording (comprising 60 input samples) for subject S1 during the presentation of stimuli 12\footnote{12 is chosen here for demonstration purposes only because it is one of the stimuli with the lowest standard deviation in terms of valence ratings provided from all subjects.}. In the temporal view, feature values at different levels are  normalized separately along the time dimension in the range $[-1, 1]$. The spatial view showcases features extracted at the snapshot at $t=29\sim 30$ seconds, which are further spatially re-normalized along all nodes to ensure uniformity. At the region level, nodes belonging to the same region are assigned with the same color as the region-level tomography plot and interpolated along regions, emphasizing larger scale/higher level spatial features. 

\begin{figure}[tbh]
    \centering
    \includegraphics[width=0.5\textwidth]{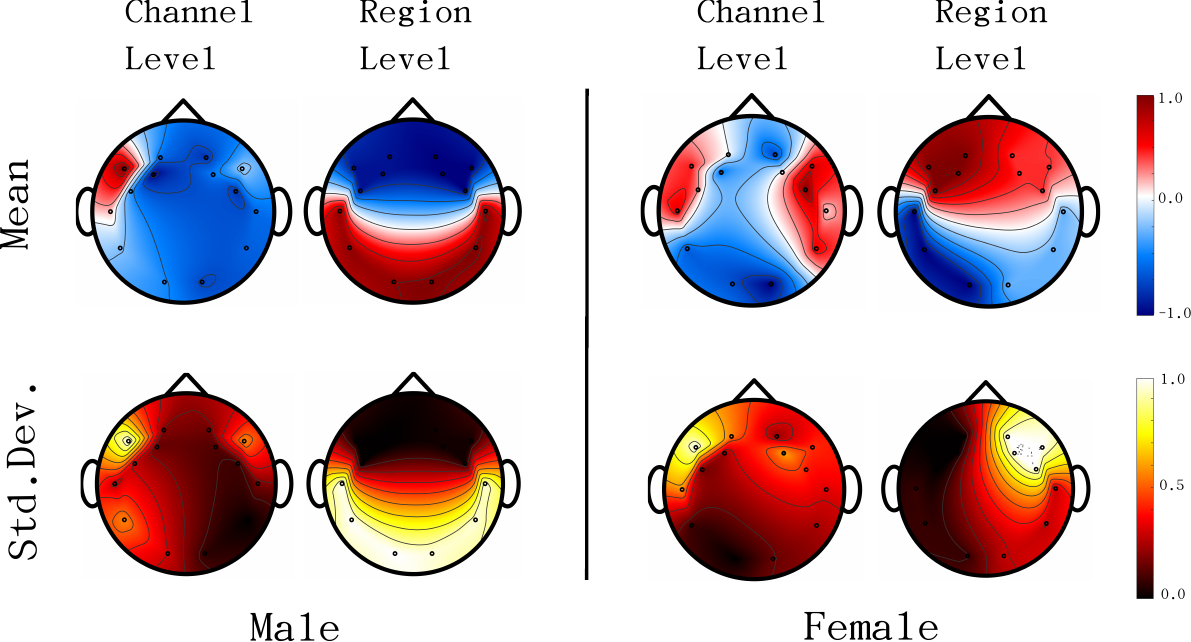}
    \caption{Channel level and region level patterns between male mean/std and female mean/std groups.}
    \label{fig:MvsF}
\end{figure}

Visualizations such as this can serve as a valuable tool to explore whether the learned patterns of the model align with clinical observations or real-world experiences. For instance, we analyze the spatial view by calculating the mean and standard deviation separately for male and female subjects, aggregating the results in Fig. \ref{fig:MvsF}. In this particular example, we observe some common patterns (e.g., high mean and standard deviation around F7 between Left Frontal/Temporal area, more complicated (pre)frontal pattern from female than male); however, there are distinct differences in feature patterns between male and female subjects. These observations appear to align with other numerical/clinical findings such as \cite{peng2023identifying,Hodgetts2023}. 

An intriguing observation is that the mean pattern at the region level between the two sexes appears to be roughly inverted. Furthermore, when examining the standard deviation pattern at the region level, we observe that the color gradients in the male pattern tend to align along the anterior-posterior direction, whereas they align along the medial-lateral direction for the female. This may indicate a greater degree of asymmetry between the left and right brain hemispheres in females. While these numerical findings may not directly correspond to clinical experiences and require further investigation, they offer an interesting representation where the logical relation of ``male-female" can be captured through simple arithmetic operations, such as taking the opposite for the mean pattern at the region level or rotating it by $90^\circ$ for the standard deviation pattern. Additionally, if one possessed strong prior knowledge regarding the region patterns, the HiSTN design allows for its enforcement during training, leading to enhanced interpretability afterwards.  

\subsection{Subject-Independent Experiments}\label{Sec:exp-si}

\begin{wrapfigure}{l}{0.25\textwidth}
  \includegraphics[width=0.25\textwidth]{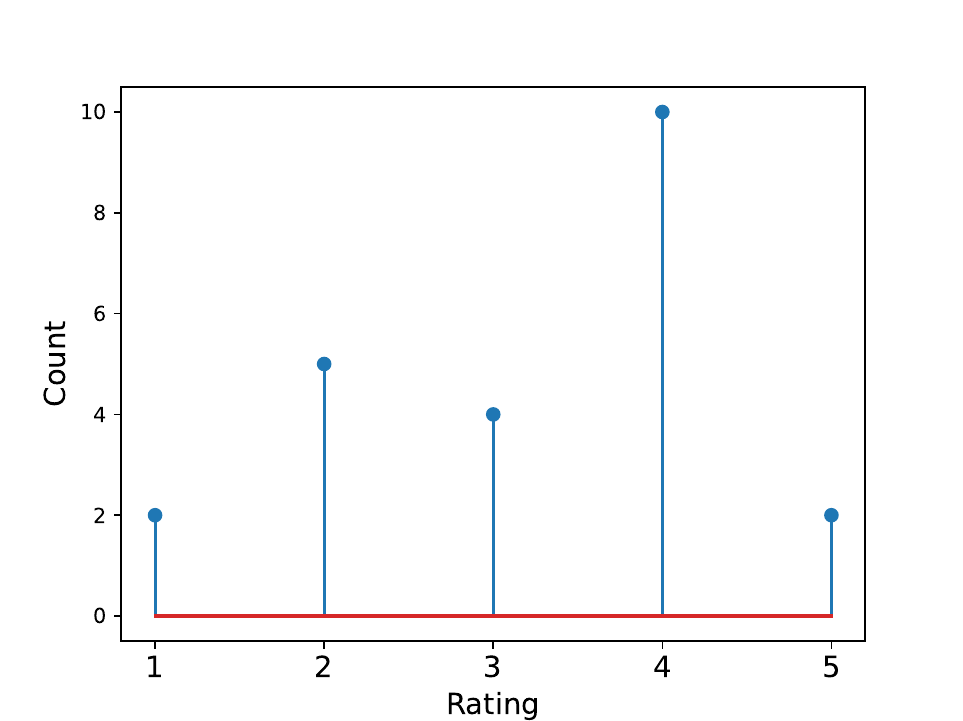}
  \caption{One example of distribution of self-reported valence ratings among all 23 subjects given the same stimuli (5th).}
  \label{fig:labeldist}
\end{wrapfigure}

Classification tasks that are independent of the subject present more significant challenges compared to those that are dependent on the subject, primarily due to the introduction of additional complexities. For instance, discrepancies often arise when different individuals provide ratings in response to the same stimuli, with some extremes being a maximum rating of 5 reported by one person and a minimum rating of 1 reported by another - an example of which can be seen in Fig. \ref{fig:labeldist}. Moreover, the interpretation of the same score, such as 3, can vary between individuals based on their distinct personalities, signifying different emotional states. In experimental scenarios like Leave-One-Out Cross Validation (LOOCV), this inconsistency in labels, given identical input from different individuals, complicates the learning of effective features by the model. 

For a more robust assessment, we continue to employ LOOCV for our experiments independent of subjects. In the case of the DREAMER dataset, the data from a single individual is chosen for testing each time, while data from the remaining subjects are combined for training purposes. While it is feasible to explore appropriate transfer learning strategies such as domain adaptation, these strategies constitute their own independent areas of research, encompass a wide range of topics and warrant considerable further investigation.

In this study, to align with our main objectives, we adopt a straightforward two-stage (pretraining/finetuning) framework to address the issue of label inconsistency across different subjects. In the \textit{first} stage, during preprocessing, for the same trial where different subjects may report varying scores, we compute the prior score distribution among different subjects. We then adjust the label to the score with the highest likelihood and employ this corrected label for training. In the \textit{second} stage, after the initial phase of training on other subjects' data, we fine-tune the model using the initial 10-second data from the target subject. We then gather the performance on 1s-segment test data drawn from the remaining 50 seconds. During this second stage, the weights contained feature head, which mainly learn low-level features, are frozen. Only blocks deeper within the network are retrained. 
These include the network blocks after depthwise convolution for both EEGNet and the proposed HiSTN, and the dense layers after the final convolution blocks in each temporal and spatial branch for MSBAM. Following this process with each subject acting as a test case once, the model's mean performance is computed and incorporated into Table \ref{tab:Si}.

\begin{table}[tbh]
    \centering
    \caption{Hyper-parameters used in benchmark for subject-independent experiments.}\label{tab:HP-SI}
    \begin{tabular}{lcc}
    \hline
        Parameter & 1st Stage & 2nd Stage \\
        \hline
        Batchsize &  120 & 100 \\
        LearningRate & $0.01$ & $0.001$\\
        Maximum Epochs & 100 & 400\\
    \hline
    \end{tabular}

\end{table}

\begin{table}[tbh]
    \centering
    \caption{Summary of subject-independent experiments on 5-classification tasks.}
    \label{tab:Si}
    \begin{adjustbox}{width=\columnwidth,center}
    \begin{tabular}{|p{5mm}|c|c|c|c|c|c|c|c|c|c|}
    \hline
    \multicolumn{3}{|c}{ } &\multicolumn{4}{|c}{Valence}  & \multicolumn{4}{|c|}{Arousal} \\
    \hline
    \multicolumn{2}{|c|}{Models} & \#Para. &F1(\%)  & Top2 Acc.(\%) & Tri-P(\%) & Seq2HR (\%) &F1(\%)  & Top2 Acc.(\%) & Tri-P(\%) & Seq2HR (\%) \\
     \hline
    \multirow{3}{3em}{\tiny{EEGNet}} 
      &  A & 3269 & \myccO 76.98 $\pm$ 5.91  &  \myccO \textbf{91.95} $\pm$ \textbf{2.62} &\myccLC 88.37 $\pm$ 4.29 & 35.68 $\pm$ 11.54 & \myccO 81.04 $\pm$ 6.98 &  \myccO \textbf{93.53} $\pm$ \textbf{3.81} &  91.99 $\pm$ 6.13 & 52.53 $\pm$ 15.88 \\ 
      &  B & 1217 & 50.26 $\pm$ 9.24  & 76.62 $\pm$ 7.56 & 86.44 $\pm$ 6.43 & \myccLC 76.62 $\pm$ 7.56 &  57.88 $\pm$ 13.36  &  83.33 $\pm$ 11.13 &  92.05 $\pm$ 7.06 &  \myccLC 82.68 $\pm$ 10.94 \\
      &  C & 8399 & 72.82 $\pm$ 8.01  &  88.25 $\pm$ 4.76 &  86.29 $\pm$ 5.11 & 48.67 $\pm$ 11.30 & \myccLC 80.14 $\pm$ 7.36   & \myccLC  93.22 $\pm$ 4.41 &  \myccLC 92.24 $\pm$ 6.93 &  60.86 $\pm$ 16.97\\
      &  D & 3269 &\myccLC 76.51 $\pm$ 6.82  &  \myccLC 89.05 $\pm$ 3.80 &  \myccO 89.65 $\pm$ 3.76 & \myccO \textbf{84.74} $\pm$ \textbf{4.89} &  77.94 $\pm$ 8.61   &   91.50 $\pm$ 6.39 &  \myccO 92.91 $\pm$ 6.74 &  \myccO 87.48 $\pm$ 9.18\\
    \hline
    \multirow{3}{3em}{\tiny{MSBAM}} 
      &  A & 203,269 & \myccO 67.22 $\pm$ 7.25   & \myccO 84.85 $\pm$ 5.05 & 81.57 $\pm$ 5.21 & 36.63 $\pm$ 14.90 & \myccO 69.64 $\pm$ 7.79   & \myccO 88.08 $\pm$ 5.47 & 86.87 $\pm$ 7.56 & 53.39 $\pm$ 18.76\\ 
      &  B & 202,785 &52.38 $\pm$ 7.22  & 74.49 $\pm$ 5.96 & \myccO 84.60 $\pm$ 4.45 & \myccLC 74.49 $\pm$ 5.96 &  54.46 $\pm$ 10.42  &  80.02 $\pm$ 9.60 & \myccO 89.71 $\pm$ 6.98 & \myccLC 79.75 $\pm$ 9.42\\
      &  C & 204,479 &65.37 $\pm$ 9.34  &\myccLC 83.54 $\pm$ 6.44 & 82.98 $\pm$ 4.44 & 52.75 $\pm$ 12.51 & 60.03 $\pm$ 14.26  &  83.01 $\pm$ 10.34 &  87.52 $\pm$ 7.65  &  63.24 $\pm$ 17.11\\
      &  D & 203,269 & \myccLC 66.51 $\pm$ 7.11  & 82.12 $\pm$ 4.78 &\myccLC 84.48 $\pm$ 4.44 &\myccO 75.53 $\pm$ 6.66 &  \myccLC 68.72 $\pm$ 8.54   & \myccLC 85.81 $\pm$ 7.34 &\myccLC 89.62 $\pm$ 6.93 &  \myccO 81.89 $\pm$ 9.53\\
    \hline
    \multirow{3}{3em}{\tiny{HiSTN}}    
      &  A & 1181 & \myccLC 77.02 $\pm$ 5.63  & \myccO 91.23 $\pm$ 3.42 &  87.49 $\pm$ 4.04 & 37.18 $\pm$ 14.45 & \myccLC 76.57 $\pm$ 8.54  & \myccLC 92.18 $\pm$ 4.71 &  89.49 $\pm$ 7.95 & 53.66 $\pm$ 18.19\\ 
      &  B &  1101 &  55.69 $\pm$ 7.46  &   79.44 $\pm$ 6.58 & \myccLC 89.09 $\pm$ 5.02 &  \myccLC 79.44 $\pm$ 6.58 &  62.63 $\pm$ 13.93   &  85.18 $\pm$ 10.80 & \myccLC 93.08 $\pm$ 6.78 &  \myccLC 84.94 $\pm$ 10.68\\
      &  C\footnote{We used learning rate 0.005 and 500 epochs at the 2nd training stage for this, because the hyperparameters as shown in Table \ref{tab:HP-SI} resulted in very poor performance indicating the model was not sufficiently trained.} 
           & 1381 &65.19 $\pm$ 6.21  & 84.91 $\pm$ 4.94 & 87.66 $\pm$ 5.00 & 70.87 $\pm$  12.36 &  67.14 $\pm$ 8.32    &  88.15 $\pm$ 6.52 & 91.47 $\pm$ 6.95 &  77.17 $\pm$  15.36\\
      &  D & 1181 & \myccO \textbf{78.34} $\pm$ \textbf{5.53} & \myccLC 90.40 $\pm$ 3.02 & \myccO \textbf{90.59} $\pm$ \textbf{3.79} & \myccO 82.61 $\pm$ 5.99 & \myccO \textbf{81.59} $\pm$ \textbf{7.06}   & \myccO 92.47 $\pm$ 5.18 &  \myccO \textbf{93.61} $\pm$ \textbf{5.47} & \myccO \textbf{88.62} $\pm$ \textbf{8.65}\\
     \hline
    \end{tabular}
\end{adjustbox}
\end{table}

Subject-independent tasks, with their notably larger train/test gap, inherently present a more complex challenge, resulting in prediction performance that does not quite match the levels observed in subject-dependent experiments. Nonetheless, certain findings noted in these two tables align with earlier subject-dependent studies. Irrespective of the specific models utilized, the proposed special label smoothing consistently elevates the Seq2HR value significantly\footnote{Similarly as seen in the subject dependent benchmark, pairwise t-test scores give very small p-values (i.e. EEGNet: $2.74e^{-16}/1.29e^{-12}$, MSBAM:$5.35e^{-13}/1.61e^{-9}$, HiSTN: $9.53e^{-14}/1.18e^{-11}$, for valence/arousal respectively, suggesting, e.g., strong statistical significance). }. Moreover, under the same setting, Variation B (the regression task) proves more challenging to optimally train compared to the other variations. The proposed HiSTN-D configuration achieves a commendable equilibrium between pure accuracy metrics and prediction continuum, utilizing the fewest parameters. The performance decline from subject-dependent tasks to subject-independent tasks is more pronounced for MSBAM compared to the other two networks. This could be attributed to the constrained volume of data available for the second stage of training, making it harder to guide larger models such as MSBAM to adapt their prediction to specific test subjects.

\subsection{Ablation Study}\label{sec:ablation}
\subsubsection{Different choices of graphs}
\begin{figure}[tbh]
    \centering
    \includegraphics[width=0.45\textwidth]{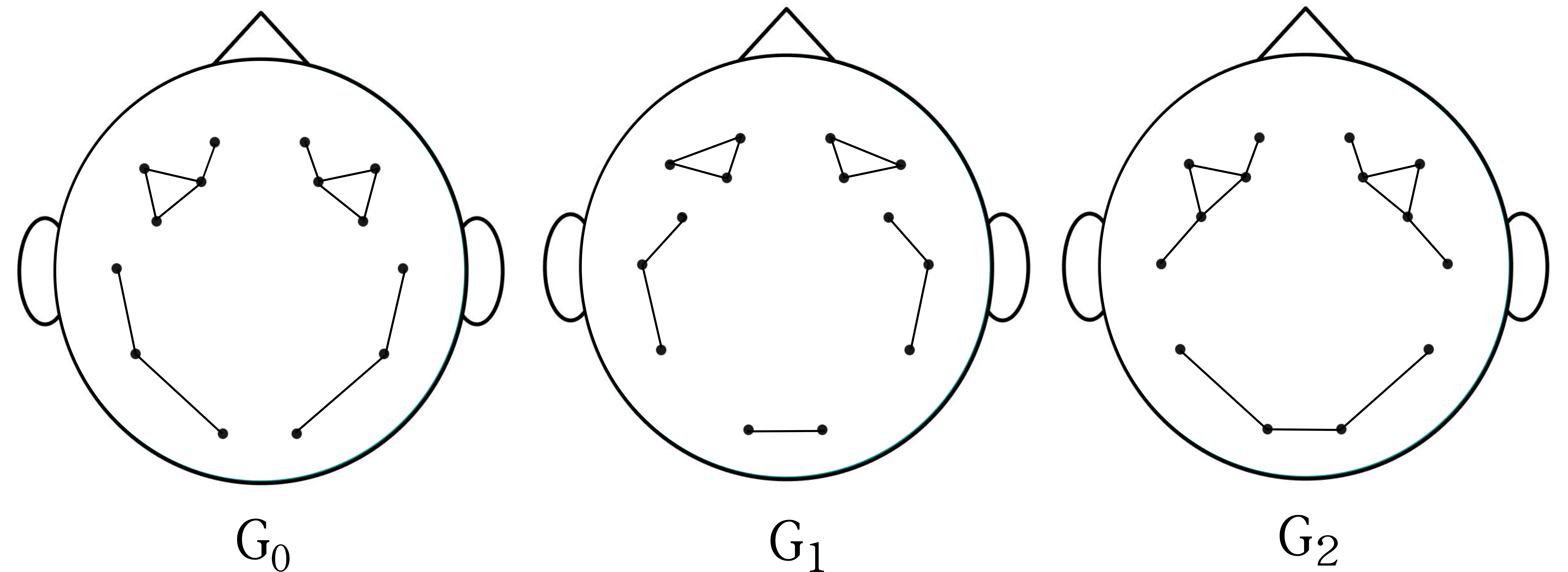}
    \caption{Different prior graph structures at the channel level. }
    \label{fig:diff_graphs}
\end{figure}

\begin{figure}[tbh]
    \centering
    \includegraphics[width=.35\textwidth,height=125pt]{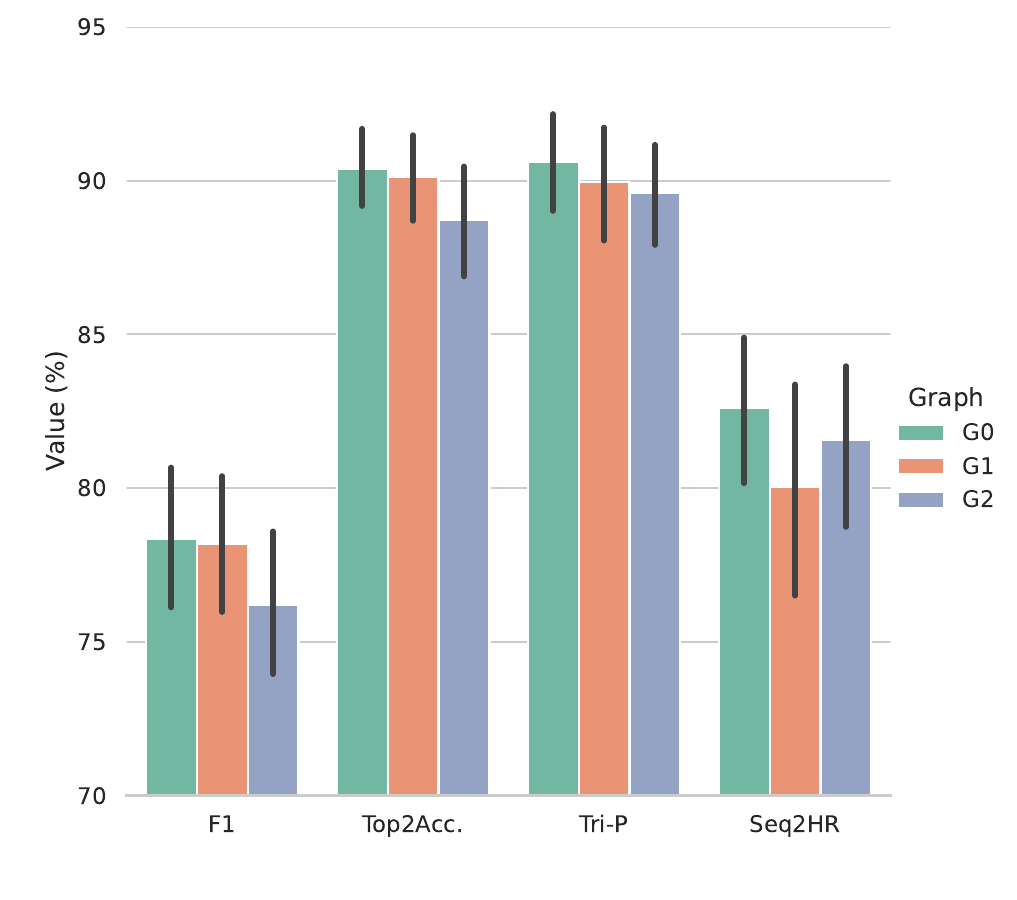}
    \includegraphics[width=.35\textwidth,height=125pt]{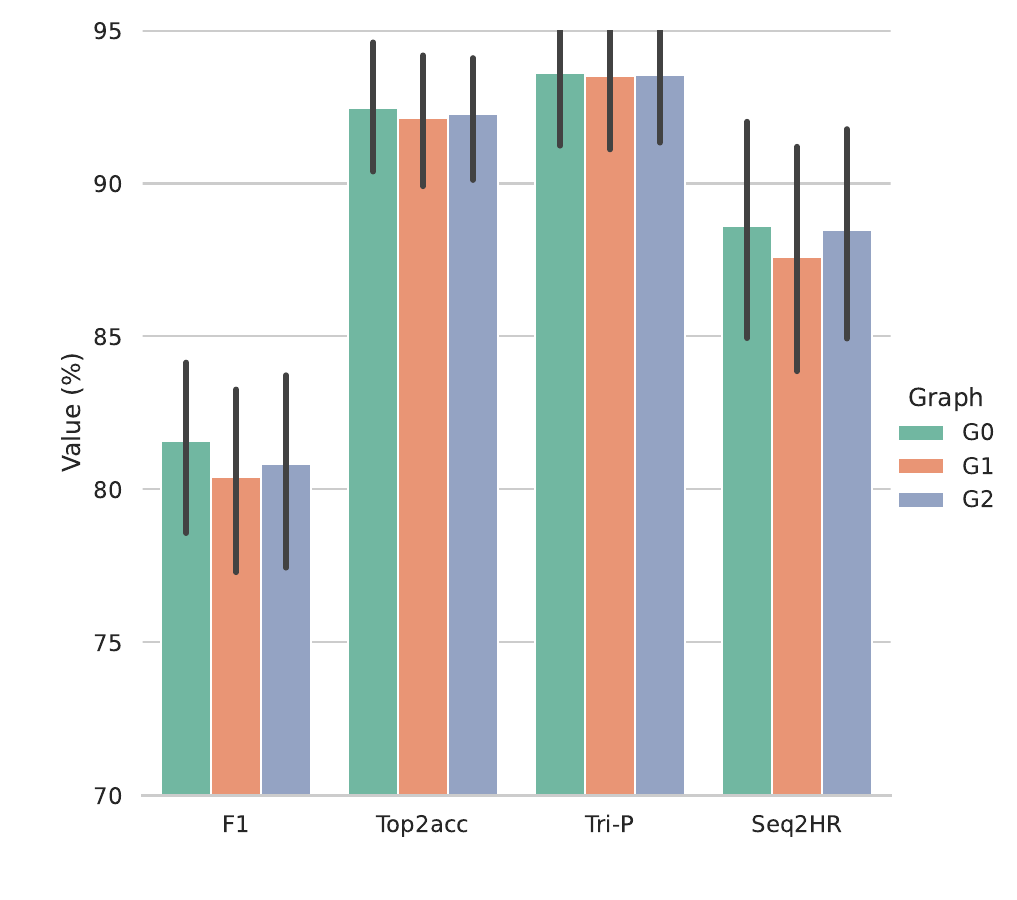}
    \caption{Performance when different prior graph structures are adopted. Left: Valence. Right: Arousal. Black lines represent the estimation of 95\% confidence intervals.}
    \label{fig:graph_ablation_m}
\end{figure}

\begin{figure}[tbh]
 \centering
 \includegraphics[width=0.5\textwidth]{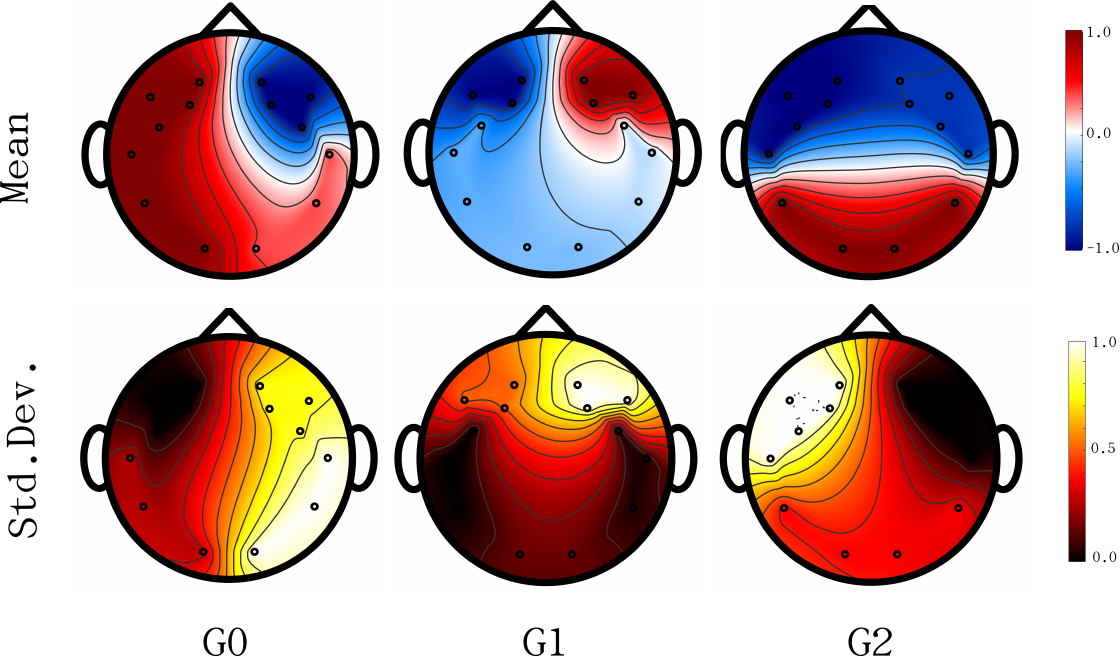}
 \caption{Patterns from the mean and standard deviation along the time direction for deep features learned at the region level when different graph structure is adopted in HiSTN. }
\label{fig:graphfeature}
\end{figure}

Fig \ref{fig:diff_graphs} presents three distinct constructions of channel-level graphs, each of which corresponds to a regional level graph - specifically a 4-cycle, a 5-cycle, and a 3-cycle graph. In accordance with the HiSTN-D configuration, we implement an identical benchmark protocol for subject-independent studies as outlined in Section \ref{Sec:exp-si}. The predictive performance for each construction is collated in Fig. \ref{fig:graph_ablation_m}. From our experimentation, the $G_0$ configuration emerged as the superior choice in terms of overall performance.
Though numerical benchmarks can be close, different graph structures employed in HiSTN can yield varied patterns of deep features, an aspect lightly touched upon at the conclusion of Section \ref{Sec:exp-sd}. For a more tangible exploration, we use data from subject 12—chosen due to high prediction performance across all three variations—on stimuli 12. We illustrate the mean and standard deviation of the spatial view on the regional level in Fig. \ref{fig:graphfeature} (where both mean and standard deviation are calculated over time). As anticipated, different deep feature patterns emerge as a consequence of the varied choices in graph structures.

\subsection{Comparing with DiffPool} \label{sec:diffP}
In Table \ref{tab:vs.DP}, we evaluate the performance of HiSTN when the graph hierarchy (part II), as illustrated in Fig. \ref{fig:node_fusion}, is replaced with DiffPool layers (designated as HiSTN(DP) in the table). This evaluation is carried out under the aforementioned subject-independent tasks, using the proposed label smoothing technique. The implementation of DiffPool layers were from Spektral \cite{Spektral}. All other experimental parameters remained consistent with those outlined in Section \ref{Sec:exp-si}. The table clearly demonstrates that a more specialized graph hierarchy design, as proposed in Section \ref{Sec:idea}, facilitates improved predictions for this task.
\begin{table}[tbh]
    \centering
    \caption{Comparison of mean test performance when graph hierarchy in HiSTN is replaced by DiffPool layers. The green numbers are p-values from the paired t-tests suggesting the improvement is of statistical significance. }
    \label{tab:vs.DP}
    \begin{adjustbox}{width=\columnwidth,center}
    \begin{tabular}{c|c|c|c|c|c}
    \hline
    \multicolumn{2}{c|}{ } &\multicolumn{2}{c}{Valence}  & \multicolumn{2}{|c}{Arousal} \\
    \hline
    Models & \# Para. & F1 (\%) & Seq2HR(\%) & F1 (\%) & Seq2HR(\%)\\
    \hline
    HiSTN(DP)-D &  8848  & 67.22 $\pm$ 5.62  & 69.07 $\pm$ 11.92 & 68.59 $\pm$ 6.00 & 78.49 $\pm$ 11.13\\
    HiSTN-D  &  \textbf{1181}  & \textbf{78.34 $\pm$ 5.53} $( {\color{darkgreen}\uparrow 1.07e^{-7}})$ & \textbf{82.61 $\pm$  5.99} $ ({\color{darkgreen} \uparrow 4.75e^{-7}})$& \textbf{81.59 $\pm$ 7.06} $({\color{darkgreen}\uparrow 7.35e^{-11}})$ & \textbf{88.62 $\pm$ 8.65} $({\color{darkgreen} \uparrow 1.05e^{-8}})$\\
    \hline
    \end{tabular}
\end{adjustbox}
\end{table}

\section{Discussion}\label{Sec:discusion}
The experiments conducted using the proposed HiSTN model have demonstrated the possibility for a lightweight yet thoughtfully constructed model to deliver effective prediction performance, even in the context of limited data. HiSTN's hierarchical spatial and temporal architecture further allows for the integration of prior knowledge, thereby further help enhance human interpretability. This is particularly applicable in discerning the potentially meaningful spatial relations among recorded EEG data or in extracting different levels of information from brain function connectivity priors. Furthermore, our benchmark results indicate that when combined with other training techniques such as the proposed special label smoothing, HiSTN is capable of achieving better balance between quantitative and qualitative prediction. However, one must be mindful of the increased computational complexity brought on by the hierarchical design, especially during inter-layer message passing and node fusion, as compared to conventional convolution. Like other graph-based neural networks, it could be subject to common issues such as `over-squashing' \cite{alon2020bottleneck} or `over-smoothing' \cite{li2018deeper}. 

This research sets a stage for further exploration in numerous fields with the potential for significant advancements in application performance. Notably, the employed graphs and hierarchical designs could be refined through a blend of clinical knowledge and mathematical tools. Theories on extending concepts from smooth manifolds such as Ricci flows to graph-like discrete structures could potentially help optimize a graph structure initially created based on clinical priors for deep learning purposes \cite{topping2021understanding}. Concurrently, the strategic implementation of efficient transfer learning techniques might enable a seamless translation of knowledge acquired from the training domain to specific target subjects. Additionally, the challenge of label inconsistency, which can be seen as a noisy label issue or fuzzy label issue, could be more effectively tackled using reinforcement/contrastive learning techniques or be examined under fuzzy set/logic framework. These approaches, nested within the semi-supervised learning framework, have the potential to better manage the problem by appropriately weighting or selectively choosing information-rich samples.

\section{Conclusion}\label{Sec:conclusion}
This paper showcases through both subject-dependent and independent experiments on the DREAMER under the finer-grained 5-classification tasks that the proposed HiSTN can offer a highly parameter-efficient solution. By integrating the proposed spatial label smoothing technique, the quality of the model's predictions can be significantly enhanced, as indicated by the high likelihood of top predictions encapsulating true labels and their proximity to each other. Despite certain limitations and potential future areas for exploration highlighted in the Discussion Section, this study serves as a promising step towards optimizing the balance between quantitative metrics and qualitative behavior in model predictions, particularly in scenarios where data is scarce and parameter efficiency is a critical factor.

\section*{S.1: Generally message passing does not commute with temporal convolutions}

Given the feature matrix $X \in \mathcal{R}^{C \times T}$ generated by stacking signals of length $T$ from $C$ different channels, we let $A$ denote the matrix multiplied from the left for message passing ($\psi$), and $W \in \mathcal{R}^{C\times k}$ be the stack of convolution kernels of length $k$. The channel-wise time convolution $*_t$ defined for 2d feature matrix is then performed as follows: 
$$\mathcal{T}(X) = X *_t W = \begin{bmatrix}
    X_{1\cdot} * W_{1\cdot}\\
    \cdots \\
    X_{C\cdot} * W_{C\cdot}
    \end{bmatrix} 
$$
where $*$ is the regular 1d convolution along time and entries $[\ ]_{i\cdot}$ (e.g. $X_{1\cdot}$) represent the entirety of the $i$th row vector. We can then calculate $\mathcal{T}\circ \mathcal{\psi} = (AX)*_t W $ and $\mathcal{\psi}\circ \mathcal{T} = A(X*_t W)$ as follows:
\begin{align}
    \mathcal{T}\circ \mathcal{\psi}(X) = (AX)*_t W & =  \begin{bmatrix}
    \sum\limits_{j=1}^C A_{1j}X_{j\cdot} \\
    \cdots \\
    \sum\limits_{j=1}^C A_{Cj}X_{j\cdot}
    \end{bmatrix} *_t W\\
    & = \begin{bmatrix}
    \sum\limits_{j=1}^C A_{1j}X_{j\cdot} * W_{1\cdot} \\
    \cdots \\
    \sum\limits_{j=1}^C A_{Cj}X_{j\cdot} * W_{C\cdot}
    \end{bmatrix}, \\
    \mathcal{\psi}\circ \mathcal{T}(X) = A(X*_t W) & =  A \begin{bmatrix}
    X_{1\cdot} * W_{1\cdot} \\
    \cdots \\
    X_{C\cdot} * W_{C\cdot}
    \end{bmatrix} \\
    & = \begin{bmatrix}
    \sum\limits_{j=1}^C A_{1j} X_{j\cdot} * W_{j\cdot} \\
    \cdots \\
    \sum\limits_{j=1}^C A_{Cj}X_{j\cdot} * W_{j\cdot}
    \end{bmatrix}.
\end{align}

Thus, in order to have $(AX) *_t W  = A(X *_t W) $, one must require that,
\begin{equation}
    \sum\limits_{j=1}^C A_{ij}X_{j\cdot} * W_{i\cdot} = \sum\limits_{j=1}^C A_{ij} X_{j\cdot} * W_{j\cdot} \, , \quad \forall i.
\end{equation}

or equivalently, 
\begin{equation}
    \sum\limits_{j=1}^C A_{ij}X_{j\cdot} * (W_{i\cdot} -  W_{j\cdot}) = 0 \, , \quad \forall i.
\end{equation}

The following example gives a straightforward  calculation.\footnote{The term \textit{convolution} referred to in the neural network setting is actually a \textit{correlation} in standard mathematics terminology, i.e the kernel is not rotated by $180^\circ$. In the example, we followed the neural network setting, but one can easily verify that the equation does not hold for either \textit{correlation} or \textit{convolution}.}  Consider the matrices:
\[
A = \begin{bmatrix} 1 & 0.5 \\ 0.5 & 1\end{bmatrix}, 
X = \begin{bmatrix} 1 & 3 & -1 & -2 \\ -1 & 2 & 1 & 0\end{bmatrix}, W = \begin{bmatrix} -1 & 2 \\ 3 & 1\end{bmatrix},
\]
so that, 
\begin{align}
    (AX)*_t W & =  \begin{bmatrix}
    0.5 & 4 & -0.5 & -2 \\ -0.5 & 3.5 & 0.5 & -1
    \end{bmatrix} *_t \begin{bmatrix} -1 & 2 \\ 3 & 1\end{bmatrix}\\
   & = \begin{bmatrix}
    7.5 & -5 & -3.5 \\ 2 & 11 & 0.5
    \end{bmatrix} 
\end{align}
while, 
\begin{align}
    A(X*_t W) & =  \begin{bmatrix}
    1 & 0.5 \\ 0.5 & 1
    \end{bmatrix} \begin{bmatrix} 5 & -5 & -3 \\ -1 & 7 & 3\end{bmatrix}\\
   & = \begin{bmatrix}
    4.5 & -1.5 & -1.5 \\ 1.5 & 4.5 & 1.5
    \end{bmatrix},
\end{align}
thus arriving with $(AX)*_t W \neq A(X*_t W)$. 

\bibliographystyle{abbrvnat}
\bibliography{refs}  






\end{document}